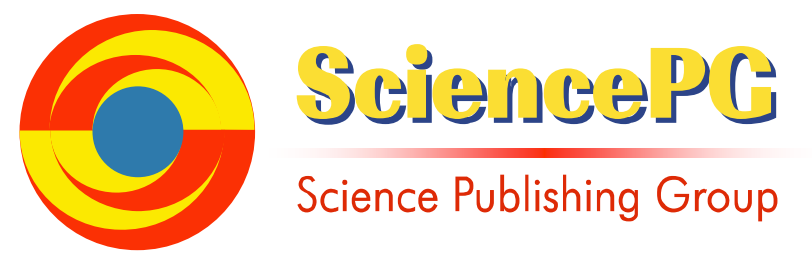

**Review Ariticle**

# A Critical Review of the Lunar Laser Ranging

**Andreas Märki**

Märki Analytics for Space, Erlenbach ZH, Switzerland

**Email address:**
a.m.maerki@bluewin.ch

**Abstract:** This paper provides an overview of the Lunar Laser Ranging (LLR) experiments. The measurement principle is explained and its theory is derived. Both contributions, the direct reflected light from retroreflectors as well as the scattered light from the lunar surface are considered. The measurement results from the Sixties until 2007 are then compared between different LLR stations and with the theoretical forecast. The very first experiment was in 1962: a laser beam was directed to the Moon and the scattered light from the lunar surface was detected. The number of received photons was in line with the theory. Then from 1969 the laser beams were directed to retroreflectors placed by Apollo astronauts and Luna space crafts. Retroreflectors are on the one hand reference points for long term measurements; on the other hand they deliver a much stronger return signal compared to the scattered return. But none of the stations could measure the expected amplification of the retroreflectors. The number of received photons remained in line with measurements to the bare surface of the Moon. Therefore either all retroreflectors have degraded such that their return signals fit to the scattered return from the lunar soil or the measurements were indeed taken to the lunar surface only.

**Keywords:** LLR, Retroreflector, Scattering

## 1. Introduction

Lunar Laser Ranging (LLR) was performed the first time in 1962, i.e. soon after the invention and the first operation of a laser in 1960. LLR then became well known by NASA's Apollo program. Since 1969 several observatories reported about their LLR experiments on lunar retroreflectors (Figure 1) providing millimetre accuracy.

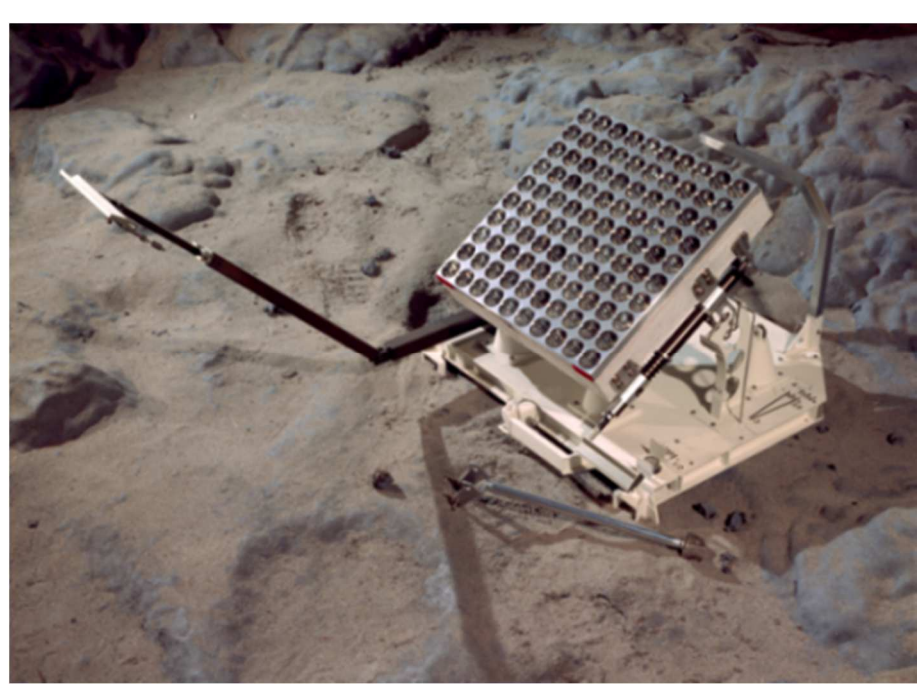

***Figure 1.** Laser Retroreflector "Apollo 11" (Science Museum).*

Often there is no or only little information about the link budget, i.e. the calculation of the number of received photons in relation to the emitted photons. Therefore I derive first the necessary theory to be able to review the different measurement results.

## 2. Measurement Principle and Applications

In the lunar laser ranging the time of flight of a laser pulse is measured. The laser pulse flies from the sender or transmitter on the Earth to the Moon and back:

$$z = c \cdot \Delta T / 2 \tag{1}$$

$z$ is the distance between the measurement station and the measuring object, i.e. the Moon; $c$ is the velocity of light, which is about 300‘000 km/s; $\Delta T$ is the time of flight of the laser pulse. The following applications operate according to the same principle:

1. Laser altimeters in aircraft, which measure the altitude

above ground.

2. Laser altimeters which measure the altitude of a spacecraft above the surface of a celestial body and so determine its ground profile.
3. Satellite Laser Ranging (SLR): measurement of the distance from a ground station to a satellite, e.g. the LAGEOS (Laser Geodynamics Satellite, Figure 2).

Another measurement principle with lasers is the interferometry. It is applied mostly for short distances where a high accuracy is required. The interferometry is not discussed further here because it is not used for the lunar laser ranging.

To increase the return signals one often uses retroreflectors. They reflect the light exactly into the direction of the incoming light.

Laser altimeters of aircraft and of spacecraft measure without retroreflectors. In the SLR all satellites to be measured are equipped with retroreflectors.

Measurements to the Moon can principally be made with or without retroreflectors. But for accurate long term measurements a retroreflector is compulsory because only then one can be sure that the measurements are done to the same reference point.

Even if the exact position of the reflector were known a direct homing is not feasible with the narrow beam needed. The reflector has to be found by a scanning motion; whether it is hit or not is recognized on the return signal: on its high signal power and also on its signature, i.e. on the small variation of the distance measurement.

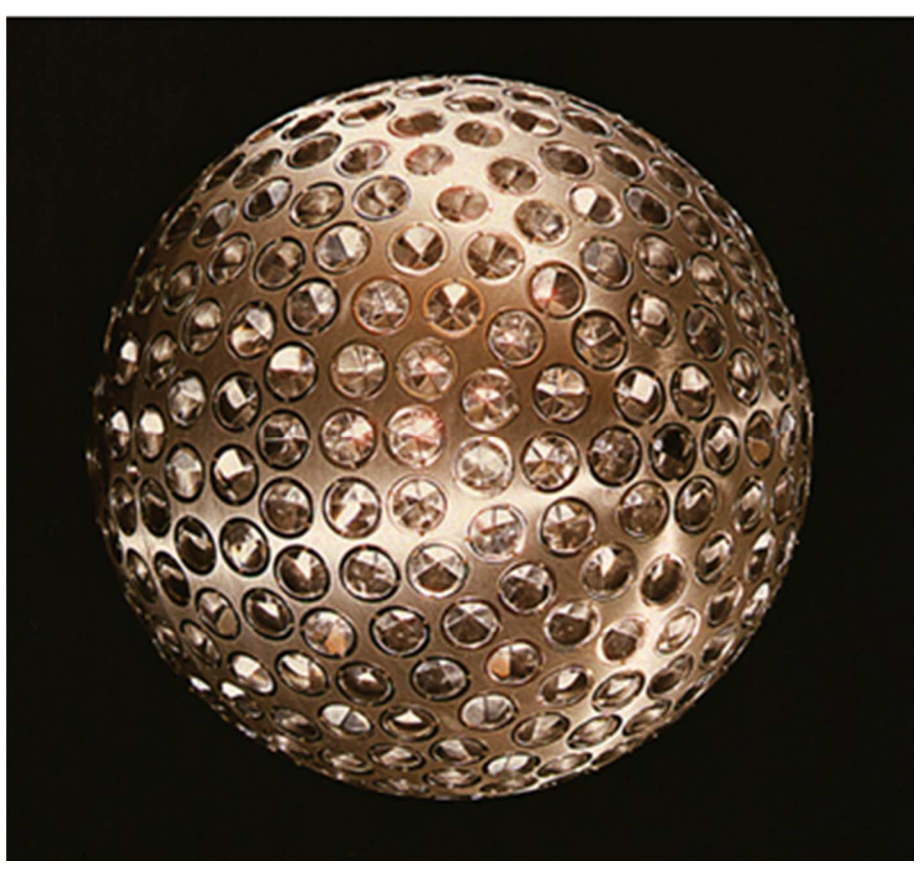

***Figure 2.*** *LAGEOS, made specifically for SLR [Wikipedia: LAGEOS].*

# 3. Theory of LLR

The emitted laser pulse must have sufficient energy or a sufficient number of photons so that at least one photon returns to the receiver.

"One photon" sounds little, but today's receivers can detect single photons with a high probability (>70%) (e.g. [1]). Since there are efficient receivers for the green light one generally selects the wave length of 532nm. This is half of 1064nm, which is the wave length of NdYAG lasers.

Here the measurements with and without retroreflectors are considered. For both the link budget is derived.

### *3.1. Link-Budget for Measurements Without Retroreflector*

In this measurement principle the laser light is scattered on the surface of the measuring object (here: surface of the Moon). The scattering is modeled as a Lambertian scattering where the light is scattered in all directions. Only a small part of the light of a surface element is therefore scattered back in the direction of the incoming light; but light is scattered back from the whole illuminated area. The field of view of the receiver is larger than the divergence of the transmit beam so that it can collect light from the whole illuminated area.

The link budget is calculated according to Thomas_2007 [2] (5) first for vacuum:

$$E_{RX} = E_{TX} \cdot T_{RX} \cdot \frac{A_{RX}}{z^2} \cdot \frac{Albedo}{\pi} \tag{2}$$

$E_{RX}$ is the received energy; $E_{TX}$ is the emitted energy; $T_{RX}$ is the transmission of the receiver telescope; $A_{RX}$ is the area of the receiver telescope; $z$ is the distance from the sender to the measuring object; *Albedo* is here the reflectivity of the measured surface for the used wavelength.

For a measurement through the atmosphere one has additionally to consider the atmospheric transmission $T_{atm}$ for the run forward and backward. If the receiver telescope has a circular entrance, $A_{RX}$ can be written as $R_{RX}^2 \cdot \pi$, if a potential central obscuration is neglected. If the emitted power is measured behind the transmit telescope then its transmission ($T_{TX}$) has to be considered as well. Additionally the quantum efficiency $\eta_q$ (≈0.7) of the detector is introduced. The adapted equation (2) looks then as follows:

$$\frac{E_{RX}}{E_{TX}} = \eta_q \cdot T_{TX} \cdot T_{RX} \cdot T_{atm}^{\ 2} \cdot \frac{R_{RX}^{\ 2}}{z^2} \cdot Albedo \tag{3}$$

As numerical values for all transmissions 0.707 is taken. These values seem on the one hand reasonable and simplify on the other hand the calculation, because $0.707^2=0.5$. For the atmospheric transmission this is valid according to Degnan_1993 [3] (figure on page 9) for quite good weather conditions. For the optical transmission through a telescope also 0.707 is assumed; this value considers mainly the central obscuration and for the transmit telescope additionally the Gaussian beam profile (Klein_Degnan_1974 [4], figure on page 4). The albedo of the Moon is ≈0.1, and the distance from the Earth to the Moon is set to 380'000km (measuring distance, not distance of the centres).

So (4) shows the link budget of the lunar laser ranging with numerical values:

$$\frac{E_{RX}}{E_{TX}} = 0.7 \cdot 0.5 \cdot 0.5 \cdot \frac{R_{RX}^{\ 2}}{(380'000km)^2} \cdot 0.1 = 0.0175 \cdot \frac{R_{RX}^{\ 2}}{(380'000km)^2} \tag{4}$$

### *3.2. Link-Budget for Measurements with Retroreflectors*

At this measurement principle the light which hits the retroreflectors is mainly reflected back along the direction of the incoming light. But the longer the measuring distance, the less light hits the retroreflector. Therefore the retroreflector must be sufficiently large so that more light is reflected back than during laser ranging to the bare surface of an object.

The link budget is set up according to Degnan_1993 [3] (3.1.1) for a homogenous atmosphere and perpendicular incidence of the light on the retroreflector:

$$n_{RX} = \eta_q \cdot n_{TX} \cdot T_{TX} \cdot G_{TX} \cdot \sigma_{RR} \cdot \left(\frac{1}{4 \cdot \pi \cdot z^2}\right)^2 \cdot A_{RX} \cdot T_{RX} \cdot T_{atm}^2 \cdot T_{Cir}^2 \quad (5)$$

$n_{RX}$ is the number of received photons; $\eta_q$ is the quantum efficiency of the detector (receiver); $n_{TX}$ is the number of emitted photons; $T_{TX}$ is the transmission of the transmit telescope; $G_{TX}$ is the antenna gain of the transmitter; $\sigma_{RR}$ is the optical cross section of a single retroreflector (cube corner) (see below); $z$ is the distance from the sender to the measuring object; $A_{RX}$ is the area of the receiver telescope; $T_{RX}$ is the transmission of the receiver telescope; $T_{atm}$ is the one way transmission through the atmosphere; $T_{Cir}$ is the one way transmission through possible cirrus clouds.

$T_{Cir}$ is set to 1, i.e. it is assumed that there are no disturbing cirrus clouds during the measurement.

$\sigma_{RR}$ is defined in Degnan_1993 [3] (6.1.1) for a single retroreflector as follows:

$$\sigma_{RR} = R_{RR} \cdot \frac{4 \cdot \pi \cdot A_{RR}}{\Omega_{RR}} = R_{RR} \cdot \frac{4 \cdot \pi \cdot A_{RR}^2}{\lambda^2} \quad (6)$$

$R_{RR}$ is the reflectivity (0.3 is assumed according to [12]: the cubes are uncoated); $A_{RR}$ is the area of a retroreflector ($=D_{RR}^2 \cdot \pi/4$); $\Omega_{RR}$ is the illuminated solid angle of a retroreflector; $\lambda$ is the wave length: here 532nm (green).

Equation (5) is expanded as follows:

1. $\sigma_{RR}$ is inserted according to (6)

2. for $G_{TX}$ the diffraction limited gain of $4 \cdot \pi \cdot A_{TX}/\lambda^2$ is inserted

3. an additional efficiency factor $\eta_{add}$ is introduced

4. it is considered, that there are several ($n_{RR}$) retroreflectors present:

$$\frac{n_{RX}}{n_{TX}} = \eta_{add} \cdot \eta_q \cdot T_{TX} \cdot \frac{4 \cdot \pi \cdot A_{TX}}{\lambda^2} \cdot n_{RR} \cdot R_{RR} \cdot \ldots$$

$$\ldots \cdot \frac{4 \cdot \pi \cdot A_{RR}^2}{\lambda^2} \cdot \left(\frac{1}{4 \cdot \pi \cdot z^2}\right)^2 \cdot A_{RX} \cdot T_{RX} \cdot T_{atm}^2 \quad (7)$$

$$\frac{n_{RX}}{n_{TX}} = \eta_{add} \cdot \eta_q \cdot T_{TX} \cdot T_{RX} \cdot T_{atm}^2 \cdot n_{RR} \cdot R_{RR} \frac{A_{TX} \cdot A_{RR}^2 \cdot A_{RX}}{\lambda^4 \cdot z^4} \quad (8)$$

Equation (8) is still too idealized because the transmit beam, which may use only a part of $A_{RX}$ (i.e. $A_{TX} < A_{RX}$) or which is intentionally expanded, could be further expanded by the (turbulent) atmosphere and so the transmit gain could be reduced. Instead of a real antenna area an antenna with a radius equal to the transverse atmospheric coherence length $\rho_0$ is selected – on the basis of Degnan_1993 [3] (3.9.9). Instead of $\rho_0$ one can often see the Fried parameter $r_0$ which is about $2 \cdot \rho_0$. At good seeing conditions $\rho_0$ is >10cm [5] (at 500 nm), but for a ground station at sea level the value is smaller, i.e. about 1 to 2cm:

$$\frac{n_{RX}}{n_{TX}} = \eta_{add} \cdot \eta_q \cdot T_{TX} \cdot T_{RX} \cdot T_{atm}^2 \cdot n_{RR} \cdot R_{RR} \cdot \frac{(\rho_0^2 \cdot \pi) \cdot A_{RR}^2 \cdot A_{RX}}{\lambda^4 \cdot z^4} \quad (9)$$

The retroreflector of Apollo 11 is an array of 100 ($n_{RR}$) single reflectors, each with a diameter ($D_{RR}$) of 3.8cm (A11_PSR_1969 [6] page 167).

For $\eta_q$ and the transmissions the same numerical values as in the previous paragraph are taken.

Equation (9) is now the basis to calculate the number of received photons. Besides the receiver aperture the following two parameters are varied:

1. $\rho_0$: Transverse atmospheric coherence length (10cm and 2cm). If $R_{TX}$ is smaller than $\rho_0$, then $R_{TX}$ replaces $\rho_0$. But 10cm correspond to a divergence of 0.7" which is still larger than the divergence used in Apache Point (<0.5").
2. $\eta_{add}$: Additional efficiency factor to consider possible neglects (1 and 0.5)

The velocity aberration is neglected because it is much smaller than the beam angle of the retroreflector (A11_PSR_1969 [6] page 167).

# 4. Expected and Measured Values

The following three Lunar Laser Ranging (LLR) stations:

1. Apache Point Observatory, 2'788m altitude, USA.

2. Wettzell, 600m altitude, Germany.

3. Observatory of the Cote d'Azur, 1'270m altitude, France as well as the estimation in Dickey_1994 [7] page 5 & 6 and the LIDAR measurements Luna See in 1962 are compared.

First the number of received photons per pulse is calculated according to the equations (4) and (9).

Table 1 shows the expected number of reflected photons – reflected on the retroreflector, and Table 2 shows the expected number of photons for a measurement without any retroreflectors, i.e. scattered on the surface of the Moon.

During a measurement onto a retroreflector the reflected as well as the scattered photons are counted. Table 3 shows the measurement results.

The measured values correspond well with the expected number of scattered photons. No amplification of the return signal by the retroreflector array could be measured.

The first three Apache Point measurements in Table 3 are apart from a factor of 2.3 in the expected range of 0.11 scattered photons. The 0.135 photons/pulse have been marked as "record returns", i.e. previous measurements had obviously been less fruitful; it is not said whether this marking was absolute or relative to the telescope size.

The "occasionally" peaks of 0.6 photons per pulse are a factor of 5.5 above the scattering budget, but at least a factor of 351) (for an altitude of 2'788 m rather a factor of 900) below of what would have been expected for the homing on the retroreflector array of Apollo 15. Since these peaks were not reproducible they have only an informal character.

A measured peak of a factor of 5.5 above the scattering-budget would be high, but it can still be within the uncertainty. Specifically the atmospheric transmission can also be better; and the albedo is not constant over the whole surface of the Moon. Such a measurement could have been made on a spot with a higher reflectivity – at least in the direction of the incoming laser beam, and additionally one might have benefited from the opposition surge, i.e. from an increase of the albedo if the illumination direction coincides with the direction of observation. This effect is based on the fact that the whole observed area is illuminated and that there are no visible shadows – contrary to a general constellation at which the albedo is determined.

1) The Apollo 15 retroreflector array consists of 300 cube corners; it is therefore 3 times larger than the one of Apollo 11 for which the budget was made. The minimum expectation is therefore 3 times higher than the minimum predicted 7 photons. 0.6 photons are measured as short time peaks: 35=3·7/0.6.

***Table 1.*** *Expected Number of Reflected Photons (Reflected on the Apollo 11-Retroreflector).*

| | **Parameter** | | **Apache Point** | **Wettzell** | **ø1m Telescope** | **Cote d'Azur** |
|---|---|---|---|---|---|---|
| | $\eta_{add}$ | $\rho_0$ | | | | |
| (RX) Telescope-ø | | | 3.5 m | 0.75 m | 1 m | 1.5 m |
| Transmitted number of photons per pulse | | | $3{\cdot}10^{17}$ | $10^{19}$ [a] | $10^{21}$ [b] | $8{\cdot}10^{17}$ [c] |
| | 1 | 0.1 m | 370 | 560 | 100'000 | 180 |
| Expected number of received photons according to (9) | 0.5 | 0.1 m | 180 | 280 | 50'000 | 90 |
| | 0.5 | 0.02 m | 7 | 11 | 2'000 | 4 [d] |

a [8]: Pulse chain consisting of 10 pules with $10^{18}$ photons each.

b Virtual example: Dickey_1994 [7] page 6 predicts a loss of $\approx 10^{-21}$; he expects therefore one single photon out of the here assumed $10^{21}$ – instead of the minimum calculated 2'000.

c [9] page 45: pulse energy of 300mJ.

d The estimation in [9] is 32 (attenuation of $4{\cdot}10^{-17}$ without atmospheric damping, page 45): Multiplying it with $T_{atm}^2$ gives 16, which is slightly above this conservative value.

Table 1 shows the number of photons as they should have been expected by the LLR stations if measured on the Apollo 11 retroreflector. From the observatories only the Cote d'Azur has published such a number, and it matches the conservative values in the table (footnote d). The estimation of Dickey_1994 [7] is more than three orders of magnitudes lower.

Two values are parameterized. $\rho_0$=0.02m is slightly worse than the strong turbulence case as addressed in Degnan_1993 [3] just after (3.9.9).

***Table 2.*** *Expected Number of Scattered Photons (Scattered on the Lunar Surface)*

| | **Apache Point** | **Wettzell** | **ø1m Telescope** | **Cote d'Azur** | **Luna See** |
|---|---|---|---|---|---|
| (RX) Telescope-ø | 3.5 m | 0.75 m | 1 m | 1.5 m | 1.22 m |
| Transmitted number of photons per pulse | $3{\cdot}10^{17}$ | $10^{19}$ | $10^{21}$ | $8{\cdot}10^{17}$ | $1.75{\cdot}10^{20}$[a] |
| Expected number of received photons according to (4) | 0.11 | 0.17 | 30[b] | 0.06 | 8 |

a Corresponding to 50 J, red light ([10] page 673) (λ=694nm).

b Note: Dickey_1994 [7] estimated 1 photon for the case of a retroreflector.

***Table 3.*** *Measured Number of Scattered and Reflected Photons*

| | **Apache Point [11]** | **Wettzell [8]** | **Cote d'Azur [9]** | **Luna See [10]** |
|---|---|---|---|---|
| (RX) Telescope-ø | 3.5 m | 0.75 m | 1.5 m | 1.22 m (48") |
| Transmitted number of photons per pulse | $3{\cdot}10^{17}$ | $10^{19}$ | $8{\cdot}10^{17}$ | $1.75{\cdot}10^{20}$ |
| Measured number of received photons per pulse | 0.1087[a]<br>0.135[b]<br>0.25[c]<br>0.6[d] | < 1 | ≈ 0.01[e] | 12[f] |

a [11] page 32, Apollo 11 array.

b [11] §8 „record returns" in October 2005, Apollo 15 array (3 times larger than Apollo 11 array).

c [11] §8 „In subsequent months… rates of 0.25 photons per shot", Apollo 15 array.

d [11] §8 „occasionally peaking", Apollo 15 array.

e [9] page 45. Remark: 0.01 photons is less than expected from the scattering. This can be caused by a (too) short range gate (i.e. measurement depth) or by a too small receiver field of view.

f [10] page 673; Note: Scattering only; 12 fits well the conservative value of 8 as calculated above.

In the measurement Murphy_2007 [11] page 32 the variation of the measured distance is very small, i.e. similar as expected for a measurement onto a retroreflector. But such an effect can also be achieved if one measures onto a surface perpendicular to the measurement direction. This is indeed possible because the beam is very narrow. According to Degnan_1993 [3] (3.9.9) its divergence φ is $\pm\lambda/(\pi{\cdot}\rho_0)$. This corresponds with $\rho_0$=2cm (φ=±8.5µrad=±1.75") to a radius on the Moon of 3.2km; with $\rho_0$=10cm the radius is only 640m and the corresponding spot area is 1.3km$^2$.

A last point has not been numerically considered yet: a flat retroreflector array is sensitive on the incident beam direction. The back reflected light drops fast if the incident beam direction is not perpendicular. This is shown in Degnan_1993 [3] (figure on page 23): at a deviation of the incident beam direction of 13° only 50% is reflected back and at a deviation of ≥40° nothing at all; A11_PSR_1969 [6] page 166 presents a worse behaviour for the described retroreflector. There is an additional drop factor of 2 because the cube corners are recessed! But since the Moon shows us always the same side retroreflectors could be aligned to the mean direction to the Earth. The variation due to the libration of the Moon and the location on the Earth is within ±12°. This means that an additional signal loss would always be smaller than a factor of 4.

## 5. Further Information

The final report after the Apollo Missions [12] describes range measurements of the McDonald Observatory. These measurements were performed with a ruby laser with red light. The theoretical amplification of the retroreflectors is estimated to a factor of 10 to 100 compared to the scattered light from the lunar surface.

In 2013 further measurements of the Apache Point Observatory are reported [13]. Besides measurements to the Apollo retroreflectors also measurements to the Lunokhod 1 and 2 retroreflectors were made. It is stated that all retroreflectors, i.e. Apollo and Lunokhod, suffer from a degradation of about a factor of 10. The indicated record returns are higher than the values in Table 3.

From 2014 to 2016 the Observatory of the Cote d'Azur made infrared (IR) measurements to the three Apollo and the two Lunokhod retroreflectors [14]. Mainly relative measurements between the different retroreflectors are reported. IR allows better measurements during new and full Moon.

Retroreflector arrays have multiple reflection points which spread the return pulse. Therefore feasibility studies with larger-size reflectors are ongoing to get a perfect return pulse from one single hollow corner cube. For such large corner cubes the velocity aberration has to be considered. This can be made by adapting the internal angles of the corner cubes [15].

## 6. Summary and Conclusion

The measurements of 4 LLR stations and data of an invited LLR review paper have been compared with the theoretical data. The very first LLR station which measured onto the surface of the Moon in 1962 presented consistent data. The other 3 LLR stations reported about measurements to lunar retroreflectors, but no reproducible amplification of the reflected laser pulse compared to a measurement onto the surface of the Moon could be demonstrated.

The only indication of a retroreflector was the signature of the return signal, i.e. its small variance. But a small variance would also appear in a measurement onto a lunar surface which is perpendicular to the measurement direction.

If retroreflectors had been hit then the degradation of all of them would have had to be such that just the number of scattered photons had resulted – or even less.

One observatory, the one of the Cote d'Azur, showed a forecast for a retroreflector measurement. It well matched the here presented theory. The actual measurement was then 1'600 times smaller (=16/0.01).

The invited LLR review paper [7] predicts a loss of $10^{-21}$. This is 2'000 times smaller than the lower end as calculated here. Even the return of a measurement onto the surface of the Moon is 30 times higher. All this, together with the measurement results, may call the following statement into question: "these retroreflector arrays… are still operating normally after 25 years" (Dickey_1994 [7] page 3).

In Murphy_2007 [11] §8 a "possible degradation of lunar reflectors" is mentioned.

According to the number of return photons I go even further and conclude that in all lunar laser ranging experiments the measurements were taken to the bare surface of the Moon.